\documentclass[twocolumn,twoside]{IEEEtran}

\ifCLASSINFOpdf

\else

\fi

\ifCLASSOPTIONcompsoc
\usepackage[caption=false, font=normalsize, labelfont=sf, textfont=sf]{subfig}
\else
\usepackage[caption=false, font=normalsize]{subfig}
\fi
\usepackage{lipsum}%
\usepackage[dvipsnames]{xcolor}

\usepackage{balance}
\usepackage{multicol}   
\usepackage{cite}
\usepackage{gensymb}
\usepackage{multirow}
\usepackage{graphics}
\usepackage{epsfig}
\usepackage{graphicx}
\usepackage{epstopdf}
\usepackage{textcomp}
\usepackage{amsmath}
\usepackage{mathtools}
\usepackage{filecontents}
\usepackage{lipsum,color}
\usepackage{amssymb}
\usepackage{float}
\usepackage{comment}

\usepackage{amsthm}

\usepackage{braket}
\usepackage{times} 
\usepackage{amsthm}  

\usepackage{amsfonts}

\theoremstyle{break}

\theoremstyle{break}
\begin{document}
\raggedbottom
\makeatletter
\def\fps@figure{!t} 
\def\fps@table{!t}   
\makeatother

\title{Learning-Driven Dual-Line Laser Scanning for Fast and Accurate LEO Satellite Positioning}

\author{Mohammad~Taghi~Dabiri,~Rula~Ammuri, 
	Mazen~Hasna,~{\it Senior Member,~IEEE}, \\
	~and~Khalid~Qaraqe,~{\it Senior Member,~IEEE}	
	
	\thanks{M.T. Dabiri and K. A. Qaraqe are with the Qatar Center for Quantum Computing, College of Science and Engineering, Hamad Bin Khalifa University, Doha, Qatar. email: (mdabiri@hbku.edu.qa; kqaraqe@hbku.edu.qa).}
	
	\thanks{M. Hasna is with the Department of Electrical Engineering, Qatar University, Doha, Qatar (e-mail: hasna@qu.edu.qa).}
}

\maketitle
\begin{abstract}
	Accurate and low-latency positioning is a key enabler for optical links with Low Earth Orbit (LEO) satellites, 
	where millisecond-level beam alignment is required to maintain reliable high-data-rate communication. 
	This paper presents a \textit{learning-driven dual-line laser scanning} framework for fast and precise satellite positioning. 
	Unlike conventional Gaussian-beam acquisition systems that rely on multiple sequential beams or mechanical steering, 
	the proposed approach employs two orthogonal line-shaped laser beams to perform structured optical scanning over the ambiguity region without any moving parts. 
	A physics-based model incorporating atmospheric attenuation, turbulence, and MRR-based reflection is developed, 
	and a data-driven neural estimator is trained to map received optical energy patterns to the satellite’s two-dimensional position. 
	Simulation results demonstrate that the learning-driven method achieves near-MAP accuracy with typical errors of 7–10~m and deterministic scanning time of 1–2~ms, 
	while conventional two-stage Gaussian-beam schemes exhibit comparable errors but random sensing durations of up to 5~ms. 
	The proposed framework therefore offers a favorable trade-off between positioning accuracy, computational complexity, and sensing latency, 
	making it a practical candidate for next-generation optical LEO tracking systems.
\end{abstract}

%

%
\IEEEpeerreviewmaketitle
\section{Introduction}
\subsection{Background}
Low Earth Orbit (LEO) satellite constellations are emerging as a key technology for global broadband connectivity, remote sensing, and integrated communication–navigation services. 
At altitudes below 2000~km, these satellites move rapidly relative to the Earth's surface, offering only a few minutes of visibility during each pass over a ground station. 
Within this short contact period, fast acquisition, precise positioning, and stable tracking are essential to maintain a reliable communication link. 
Conventional radio-frequency (RF) tracking systems, however, cannot deliver the sub-meter angular accuracy and millisecond responsiveness required for next-generation optical links.

Optical wireless communication (OWC) provides extremely high data rates and immunity to spectrum congestion, but its narrow beam divergence demands highly accurate alignment. 
Even a few microradians of pointing error can disrupt the optical path, making real-time sensing and positioning indispensable for link stability. 
Thus, achieving millisecond-level and high-precision localization of LEO satellites is a fundamental requirement for practical optical downlinks.
To meet this need, this work develops a high-speed and high-accuracy positioning framework based on a dual-line laser scanning concept. 
By combining structured optical sensing with data-driven inference, the proposed approach achieves deterministic millisecond-scale scanning and near-optimal positioning accuracy, supporting future MRR-based optical LEO networks.

\subsection{Literature Review}
\textbf{Passive retroreflective sensing/positioning.}
Foundational work on retroreflectors and arrays (used for ranging and precise return geometry) underpins passive optical sensing and localization with minimal terminal complexity~\cite{Degnan2023}. Experimental retro-reflective links with fine tracking validate feasibility and characterize return statistics in mobile scenarios~\cite{Trinh2021,Quintana2021}, while all-optical retro-modulation provides a compact mechanism to embed data on the return beam~\cite{Born2018}. Recent analyses of passive retro-reflector tracking with pointing errors quantify the sensitivity of localization to beam misalignment and turbulence, motivating tighter sensing policies and better angular control~\cite{Moon2023}. 

\textbf{Acquisition and beam-tracking under photon/pointing limits.}
Adaptive acquisition for photon-limited FSO, optimal allocation between tracking and detection, and angle-of-arrival estimation for narrow beams on mobile platforms have been addressed to reduce time-to-lock and improve robustness~\cite{Bashir2020,Bashir2021,Tsai2023}. Programmatic demonstrations such as C3PO outline practical pathways to retroreflector-enabled satellite links and highlight engineering constraints that couple scan policy, aperture, and modulation hardware~\cite{C3PO2017}. Related UAV–ground MRR links quantify geometric loss and the role of array size/divergence in real deployments~\cite{DabiriTWC2022,DabiriWCL2023}.

\textbf{MRR-based satellite sensing, positioning, and joint tracking.}
Two recent studies establish a rigorous end-to-end framework for MRR-based satellite downlinks. The first develops a comprehensive model for acquisition, sensing, and positioning; derives ML-type estimators from received power; and optimizes beamwidth and sensing time under gimbal/FSM constraints and turbulence~\cite{10553228}. The second extends to joint communications-and-tracking, deriving a capacity characterization (including a closed-form lower bound) and showing how tracking error, beam geometry, and array correlation jointly govern performance; it also proposes a single-transmitter design that unifies tracking and data~\cite{10681506}. In parallel, CubeSat-enabled FSO with fine beam tracking reports joint data/beam-control strategies that move toward millisecond-level updates and high throughput (with active terminals and Gaussian beams), underscoring the value of fast, accurate positioning for LEO links~\cite{SafiTVT2025}.

\textbf{Literature gap and motivation for this work.}
Across the above body of work, sensing and positioning for satellite links are predominantly realized with Gaussian beams and multi-beam layouts (often three to five beams) that incur random acquisition/positioning duration due to scanning uncertainty. The MRR-satellite studies~\cite{10553228,10681506} rigorously quantify trade-offs (beamwidth vs. pointing loss; tracking error vs. capacity) but still assume Gaussian scans and do not provide a deterministic millisecond-level sensing mechanism with a reduced beam budget together with a learned inverse mapping from energy sequences to position. Structured line-laser illumination has not been leveraged for LEO satellite positioning in combination with an explicit learning-based estimator. This paper fills that gap by replacing randomized multi-beam Gaussian acquisition with two orthogonal line beams and a lightweight regressor, yielding deterministic \(T_{\mathrm{scan}}\approx 1\text{--}2~\mathrm{ms}\) and near-MAP accuracy while preserving the passive-terminal advantages of MRR-based links.

\subsection{Contributions}
Motivated by the limitations of existing studies, where satellite sensing and positioning rely on multi-beam Gaussian scanning with random acquisition time and several active beams, this work proposes a deterministic and low-complexity positioning framework for MRR-based LEO optical links. 
To this end, a dual-line laser scanning architecture is introduced that employs two orthogonal line-shaped beams to cover the ambiguity region without mechanical steering. 
A physics-based simulator integrating Beer–Lambert atmospheric attenuation, Gamma–Gamma turbulence, and MRR reflection models is developed to generate realistic energy–position datasets for supervised learning. 
The nonlinear relationship between the reflected optical energy sequence and the satellite position is approximated using a lightweight neural regressor trained to infer coordinates directly from the received power patterns with near-MAP accuracy.

Extensive simulations confirm the effectiveness of the proposed method. 
Compared with the classical MAP estimator, the learning-based approach achieves similar accuracy (5–10 m error range) with negligible computational cost and deterministic scanning time of 1–2 ms. 
In contrast, the conventional two-stage Gaussian-beam method provides comparable accuracy but random sensing durations between 0.5 and 3.5 ms. 
Parametric analysis further shows that an optimal beam width of about 50 m minimizes the positioning MSE to approximately 5.7² m², while extending the scan time beyond 2 ms yields little additional improvement. 
Overall, the proposed dual-line learning-based framework achieves an excellent trade-off among accuracy, latency, and complexity, offering a practical solution for next-generation LEO optical sensing and positioning systems.

\begin{figure}
	\centering
	\subfloat[] {\includegraphics[width=1.6 in]{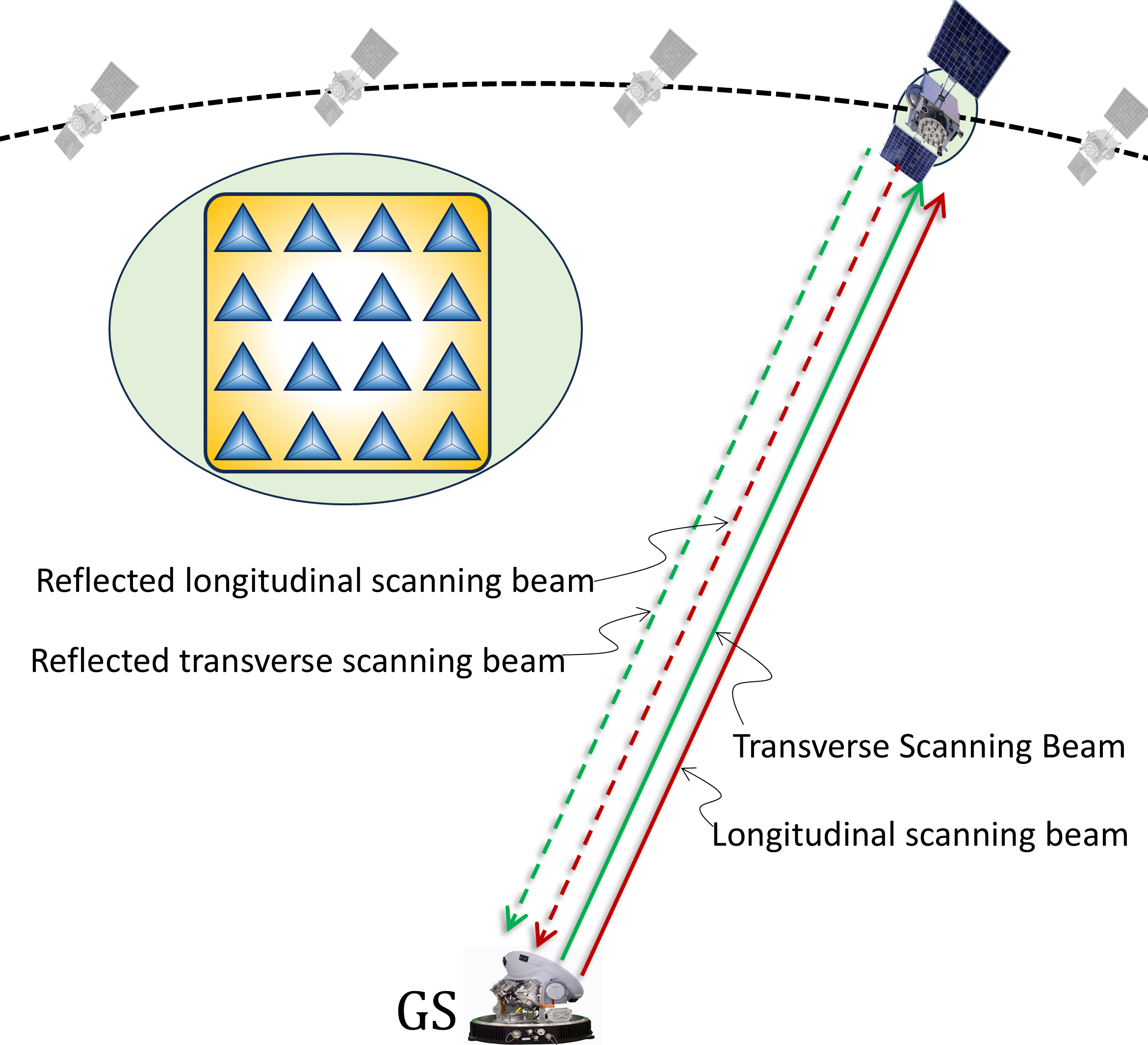}
		\label{cz1}
	}
	\hfill
	\subfloat[] {\includegraphics[width=1.6 in]{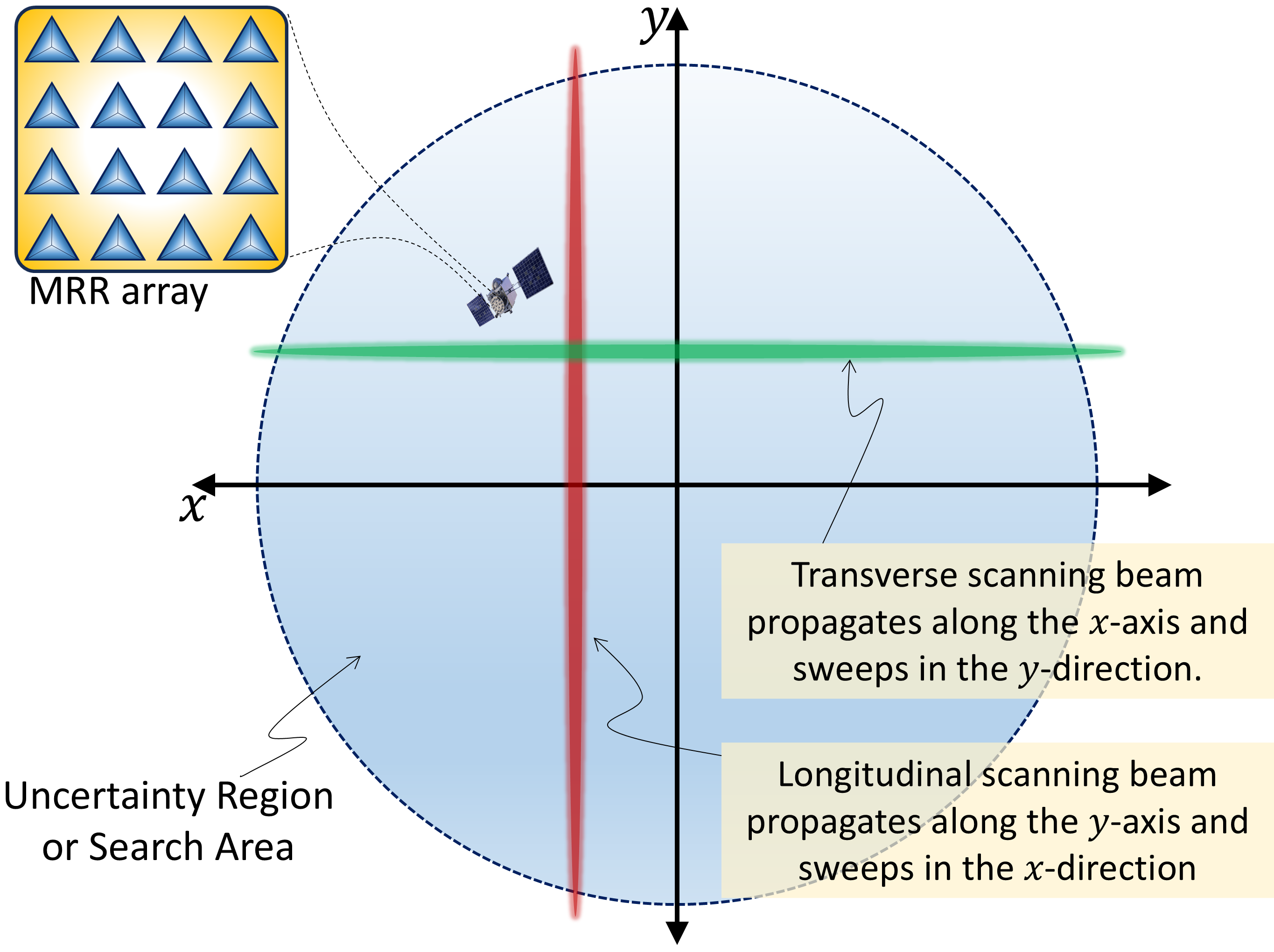}
		\label{cz2}
	}
	\caption{(a) Three-dimensional system model of the proposed dual-line laser scanning framework. 
		The longitudinal (red) and transverse (green) beams are emitted from the ground station, forming orthogonal scanning patterns approximately confined to the \textbf{$y$–$z$} and \textbf{$x$–$z$} planes, respectively.
		The satellite’s MRR array retroreflects the incident beams for bidirectional sensing and communication. 
		(b) Two-dimensional representation of the initial ambiguity (uncertainty) region on the $x$–$y$ plane. 
		The longitudinal beam sweeps in the $x$-direction, and the transverse beam sweeps in the $y$-direction to cover the search area of radius $R_{\mathrm{u}}$.}

	\label{cz}
\end{figure}
\section{System Model}\label{sec:system_model}

\subsection{System Overview and Scanning Scenario}\label{sec:overview}
As illustrated in Fig.~\ref{cz}, the proposed system consists of a ground station (GS) equipped with a dual-line laser transmitter and a high-sensitivity optical receiver, and a LEO satellite equipped with a planar array of passive optical elements. 
The GS projects two orthogonal structured laser beams, namely a longitudinal scanning beam (red) and a transverse scanning beam (green), which together form a dynamic two-dimensional illumination field.
As shown in Fig.~\ref{cz1}, the longitudinal beam propagates along the $y$-axis and performs angular sweeping in the $x$-direction, generating a scanning pattern approximately confined to the $y$–$z$ plane. 
Conversely, the transverse beam propagates along the $x$-axis and sweeps in the $y$-direction, producing a complementary pattern in the $x$–$z$ plane. 

The LEO satellite carries a modulating retroreflector (MRR) array with an effective aperture of 
$A_{\mathrm{MRR}} = N_x N_y a_{\mathrm{cell}}^2$, where $a_{\mathrm{cell}}$ is the lateral dimension of each microcell and $(N_x, N_y)$ denote the number of array elements. 
This array reflects and modulates the incident optical beams, establishing a bidirectional optical link that supports both sensing and communication functions.

At the beginning of each visibility window, when the satellite first enters the GS line-of-sight, a precise position lock is not yet available. The GS maintains only a coarse estimate of the satellite’s direction, which gives rise to an \textit{initial ambiguity region} of radius $R_{\mathrm{u}}$ on the local $x$–$y$ plane, as illustrated in Fig.~\ref{cz2}. This region represents the angular uncertainty area within which the satellite, located at position vector $\mathbf{P}_{\mathrm{sat}} = [x_{\mathrm{sat}},\, y_{\mathrm{sat}},\, z_{\mathrm{sat}}]^{\mathrm{T}}$, is expected to appear at the beginning of each visibility window~\cite{10553228,10681506}.

\subsection{Beam Model over the Search Plane}\label{sec:beam_model}
Let the two orthogonal scanning beams be indexed by $q \in \{L, T\}$, where $q=L$ and $q=T$ correspond to the longitudinal and transverse beams, respectively. 
Each beam forms an anisotropic illumination pattern on the local search plane ($x$–$y$ plane), determined by the optical divergence and the link distance between the ground station and the satellite. 
Denoting the link length by $Z_{\!q}$, the beam divergence is adjusted so that the projected footprint of each pattern spans approximately the entire acquisition region of radius $R_{\mathrm{u}}$. 
Hence, the effective beamwidths along the two in-plane axes are given by
\begin{align}
	W_{x,q} \approx 2Z_{\!q}\tan\!\Big(\frac{\theta_{x,q}}{2}\Big), 
	\quad
	W_{y,q} \approx 2Z_{\!q}\tan\!\Big(\frac{\theta_{y,q}}{2}\Big),
	\label{eq:beam_widths}
\end{align}
where $\theta_{x,q}$ and $\theta_{y,q}$ are the divergence angles along the longitudinal and transverse directions, respectively. 
To ensure full coverage of the acquisition region, the divergences satisfy
\begin{align}
	W_{x,q} \ge 2R_{\mathrm{u}}, 
	\quad 
	W_{y,q} \ge 2R_{\mathrm{u}}.
	\label{eq:coverage_condition}
\end{align}

The normalized intensity distribution of each beam $q$ on the $x$–$y$ search plane is modeled by an anisotropic super-Gaussian function \cite{parent1992propagation}:
\begin{align}
	&I_q(x,y) =\frac{1}{\pi\,w_{x,q}\,w_{y,q}} \nonumber \\
	&~~~\times \exp\!\left[-2\!\left(
	\left|\frac{x-x_{c,q}}{w_{x,q}}\right|^{2n_{x,q}} +
	\left|\frac{y-y_{c,q}}{w_{y,q}}\right|^{2n_{y,q}}
	\right)\right],
	\label{eq:beam_pattern_xy}
\end{align}
where $(x_{c,q},y_{c,q})$ denote the instantaneous beam-center coordinates on the plane, and $n_{x,q}$, $n_{y,q}$ control the flatness of the beam profile.  
For the longitudinal beam ($q=L$), the pattern is elongated along the $y$-axis and narrow along $x$, i.e.,
\begin{align}
	n_{y,L} \gg 1, \qquad n_{x,L}=1,
	\label{eq:longitudinal_shape}
\end{align}
while for the transverse beam ($q=T$) the profile is rotated by $90^{\circ}$,
\begin{align}
	n_{x,T} \gg 1, \qquad n_{y,T}=1.
	\label{eq:transverse_shape}
\end{align}

\subsection{Scanning and Timing}\label{sec:scanning_law}
During each scan period $T_{\mathrm{scan}}$, the two orthogonal beams sequentially sweep the acquisition region on the $x$–$y$ plane. 
The scanning process is discretized into a finite number of subintervals, each associated with a fixed beam orientation and duration $\tau_{q,j}$, where $j=1,2,\ldots,N_q$ and $\sum_{j=1}^{N_q}\tau_{q,j}=T_{\mathrm{scan}}$.

For the longitudinal beam ($q=L$), the beam propagates along the $y$-axis and performs angular sweeping in the $x$-direction. 
Hence, the beam center on the search plane follows
\begin{align}
	x_{c,L,j} \in [-R_{\mathrm{u}},\, R_{\mathrm{u}}], \qquad
	y_{c,L,j} = 0,
	\label{eq:center_longitudinal}
\end{align}
where $x_{c,L,j}$ denotes the instantaneous beam-center coordinate during the $j$th subinterval of the scan.
The beam moves from $-R_{\mathrm{u}}$ to $R_{\mathrm{u}}$ in $N_L$ uniform or nonuniform steps depending on the scanning strategy.
Conversely, for the transverse beam ($q=T$), the propagation axis is along the $x$-direction and the angular sweep occurs in the $y$-direction. 
Accordingly, its center coordinates are given by
\begin{align}
	x_{c,T,j} = 0, \qquad
	y_{c,T,j} \in [-R_{\mathrm{u}},\, R_{\mathrm{u}}],
	\label{eq:center_transverse}
\end{align}
where $y_{c,T,j}$ denotes the beam-center position for the $j$th subinterval of the transverse sweep.

\subsection{Reflected Signal Power Model}\label{sec:reflected_power}
Based on the analytical MRR-based optical channel model in~\cite{10681506},  
the received optical power corresponding to beam $q$ during the $j$th scanning subinterval can be expressed as
\begin{align}
	P_{r,q,j} = \eta_{\mathrm{MRR}}\,P_{t,q}\,
	\sum_{m=1}^{M} h_{q,j,m},
	\label{eq:Pr_sum}
\end{align}
where $P_{t,q}$ denotes the transmitted optical power, $\eta_{\mathrm{MRR}}$ is the overall retroreflector efficiency, 
and $h_{q,j,m}$ represents the round-trip channel coefficient associated with the $m$th MRR cell.  
Following~\cite{10681506}, each element’s coefficient can be decomposed as
\begin{align}
	h_{q,j,m}=h_{L1,q}\,h_{L2,q}\,h_{a1,q,m}\,h_{a2,q,m}\,h_{pt,q,j}\,h_{pg,q},
	\label{eq:h_components}
\end{align}
where
$h_{L1,q}$ and $h_{L2,q}$ denote atmospheric attenuation in the uplink and downlink, typically modeled by the Beer–Lambert law \cite{AndrewsBook}
\begin{align}
	h_{L1,q}=h_{L2,q}=\exp(-\zeta_{\!a}\,Z_{\!q}),
	\label{eq:beer_lambert}
\end{align}
with $\zeta_{\!a}$ as the atmospheric extinction coefficient and $Z_{\!q}$ the link distance.
$h_{a1,q,m}$ and $h_{a2,q,m}$ are independent random variables representing turbulence-induced fading for the uplink and downlink,  
each following a Gamma–Gamma (GG) distribution \cite{AndrewsBook}
\begin{align}
	f_{GG}(h;\alpha,\beta)=
	\frac{2(\alpha\beta)^{\frac{\alpha+\beta}{2}}}
	{\Gamma(\alpha)\Gamma(\beta)}\,h^{\frac{\alpha+\beta}{2}-1}
	K_{\alpha-\beta}\!\!\left(2\sqrt{\alpha\beta h}\right),
	\label{eq:gg_pdf}
\end{align}
where $\Gamma(\cdot)$ is the Gamma function, $K_{\nu}(\cdot)$ is the modified Bessel function of the second kind, 
and $\alpha,\beta$ depend on the Rytov variance determined by the turbulence level.  

The geometrical pointing term $h_{pt,q,j}$ quantifies the overlap between the beam footprint and the MRR array and will be modeled in the next subsection. 
The receiver capture factor is approximated as 
\begin{align}
	h_{pg,q} \approx \frac{4d_{\mathrm{r}}^{\,2}}{Z_{\!q}^{\,2}\theta_{\mathrm{div},q}^{2}},
	\label{eq:hpg_simple}
\end{align}
where $d_{\mathrm{r}}$ is the receiver aperture radius and $\theta_{\mathrm{div},q}$ is the divergence angle of the return beam.

\subsection{Geometrical Pointing Coefficient}\label{sec:hpt_model}
The geometrical pointing coefficient $h_{pt,q,j}$ represents the fraction of the transmitted optical power incident on the satellite’s MRR array during the $j$th subinterval. 
For the super-Gaussian beam pattern defined in~\eqref{eq:beam_pattern_xy}, the instantaneous irradiance at the satellite plane can be expressed as $I_q(x,y)$, 
where the beam center $(x_{c,q,j},y_{c,q,j})$ follows the scanning law in~\eqref{eq:center_longitudinal}–\eqref{eq:center_transverse}. 
Considering the MRR array occupies a finite aperture area $A_{\mathrm{MRR}}$ centered at the satellite position $\mathbf{P}_{\mathrm{sat}}=[x_{\mathrm{sat}},y_{\mathrm{sat}},z_{\mathrm{sat}}]^{\mathrm{T}}$, 
$h_{pt,q,j}$ is given by
\begin{align}
	h_{pt,q,j} 
	&= \frac{1}{\pi\,w_{x,q}\,w_{y,q}}
	\iint_{A_{\mathrm{MRR}}} \!\!\!
	\exp\!\Bigg[-2\!\Bigg(
	\Big|\tfrac{(x_{\mathrm{sat}}+x)-x_{c,q,j}}{w_{x,q}}\Big|^{2n_{x,q}} \nonumber \\
	&~~~+\Big|\tfrac{(y_{\mathrm{sat}}+y)-y_{c,q,j}}{w_{y,q}}\Big|^{2n_{y,q}}
	\Bigg)\Bigg]\,dx\,dy.
	\label{eq:Pintegral_explicit}
\end{align}
Since the beam footprint on the satellite plane is much larger than the MRR array aperture, 
the irradiance across $A_{\mathrm{MRR}}$ can be assumed nearly uniform. 
Under this assumption, the integral in~\eqref{eq:Pintegral_explicit} simplifies to
\begin{align}
	h_{pt,q,j} 
	&\;\;\approx\;\;
	\frac{A_{\mathrm{MRR}}}{\pi\,w_{x,q}\,w_{y,q}}\,
	\exp\!\Bigg[-2\!\Bigg(
	\Big|\tfrac{x_{\mathrm{sat}}-x_{c,q,j}}{w_{x,q}}\Big|^{2n_{x,q}} \nonumber \\
	& +\Big|\tfrac{y_{\mathrm{sat}}-y_{c,q,j}}{w_{y,q}}\Big|^{2n_{y,q}}
	\Bigg)\Bigg].
	\label{eq:hpt_approx_final}
\end{align}

As seen in~\eqref{eq:hpt_approx_final}, the received optical power in~\eqref{eq:Pr_sum} inherently embeds spatial information through the geometrical term $h_{pt,q,j}$. 
This coefficient depends on the instantaneous beam–satellite offset $(x_{\mathrm{sat}}-x_{c,q,j},\,y_{\mathrm{sat}}-y_{c,q,j})$, which varies with both the satellite’s trajectory and the scanning state $j$. 
Consequently, the temporal sequence of reflected powers $\{P_{r,q,j}\}$ carries a nonlinear and high-dimensional signature of the satellite’s position within the acquisition region. 
Because this relationship involves coupled beam dynamics, atmospheric fading, and scan geometry, an explicit analytical inversion is intractable. 
Therefore, in this work, a deep learning model is employed to learn the complex mapping from the measured power patterns to the satellite’s angular coordinates, enabling fast and accurate position estimation during each scanning cycle.

\section{Position Estimation Problem Formulation}\label{sec:problem_formulation}
The objective of the ground station is to estimate the instantaneous two-dimensional angular position of the satellite, denoted by $\mathbf{x}_{\mathrm{sat}} = [x_{\mathrm{sat}},\, y_{\mathrm{sat}}]^{\mathrm{T}}$, from the sequence of received optical powers generated by the dual-line scanning process. 
Let $\mathbf{P}_{L}$ and $\mathbf{P}_{T}$ represent the vectors of reflected powers for the longitudinal and transverse beams, respectively, defined as
\begin{align}
	\mathbf{P}_{L} &= [P_{r,L,1},\,P_{r,L,2},\,\ldots,\,P_{r,L,N_L}]^{\mathrm{T}}, \label{eq:PL_vec}\\
	\mathbf{P}_{T} &= [P_{r,T,1},\,P_{r,T,2},\,\ldots,\,P_{r,T,N_T}]^{\mathrm{T}}. \label{eq:PT_vec}
\end{align}
The overall observation vector is then
\begin{align}
	\mathbf{P} = [\mathbf{P}_{L}^{\mathrm{T}},\,\mathbf{P}_{T}^{\mathrm{T}}]^{\mathrm{T}} \in \mathbb{R}^{N_L+N_T}.
	\label{eq:P_vec_total}
\end{align}

Each measurement $P_{r,q,j}$ follows the nonlinear model in~\eqref{eq:Pr_sum}–\eqref{eq:hpt_approx_final}, compactly expressed as
\begin{align}
	P_{r,q,j} = \mathcal{F}_{q,j}(\mathbf{x}_{\mathrm{sat}}) + n_{q,j},
	\label{eq:obs_model_single}
\end{align}
where $\mathcal{F}_{q,j}(\cdot)$ encapsulates the beam geometry, atmospheric turbulence, atmospheric attenuation, and MRR reflection characteristics, and $n_{q,j}$ represents measurement noise due to detector and background noises. 
Collecting all observations, the compact measurement model becomes
\begin{align}
	\mathbf{P} = \boldsymbol{\mathcal{F}}(\mathbf{x}_{\mathrm{sat}}) + \mathbf{n},
	\label{eq:obs_model_vector}
\end{align}
where $\boldsymbol{\mathcal{F}}(\cdot) = [\mathcal{F}_{L,1},\ldots,\mathcal{F}_{L,N_L},\mathcal{F}_{T,1},\ldots,\mathcal{F}_{T,N_T}]^{\mathrm{T}}$ 
and $\mathbf{n} \sim \mathcal{N}(\mathbf{0},\,\mathbf{\Sigma}_{n})$.

\subsection{MAP-Based Estimation}
Let $\boldsymbol{\psi}_{\mathrm{sat}} = [x_{\mathrm{sat}},\,y_{\mathrm{sat}}]^{\mathrm{T}}$ denote the two-dimensional position of the satellite projected onto the local $x$–$y$ plane. 
Given the nonlinear observation model in~\eqref{eq:obs_model_vector}, the received measurement vector can be expressed as
\begin{align}
	\mathbf{P} = \boldsymbol{\mathcal{F}}(\boldsymbol{\psi}_{\mathrm{sat}}) + \mathbf{n},
	\label{eq:map_obs_compact}
\end{align}
where $\boldsymbol{\mathcal{F}}(\cdot)$ represents the nonlinear mapping induced by the beam geometry, atmospheric turbulence and attenuation, and MRR reflection, while $\mathbf{n}$ denotes additive observation noise.

The joint estimation of $\boldsymbol{\psi}_{\mathrm{sat}}$ is formulated under the Bayesian framework as the Maximum \emph{A Posteriori} (MAP) problem:
\begin{align}
	\hat{\boldsymbol{\psi}}_{\mathrm{sat}}
	&= \arg\max_{\boldsymbol{\psi}}\, p(\boldsymbol{\psi}\,|\,\mathbf{P})
	= \arg\max_{\boldsymbol{\psi}}\, p(\mathbf{P}\,|\,\boldsymbol{\psi})\,p(\boldsymbol{\psi}),
	\label{eq:map_general_compact}
\end{align}
where $p(\mathbf{P}\,|\,\boldsymbol{\psi})$ is the likelihood function governed by~\eqref{eq:map_obs_compact} and $p(\boldsymbol{\psi})$ is the prior distribution of the satellite position within the uncertainty region of radius $R_{\mathrm{u}}$.

Assuming additive white Gaussian noise (AWGN) with covariance $\sigma_{n}^{2}\mathbf{I}$, the likelihood function takes the form
\begin{align}
	p(\mathbf{P}\,|\,\boldsymbol{\psi})
	\propto 
	\exp\!\left(
	-\frac{1}{2\sigma_{n}^{2}}
	\big\|
	\mathbf{P}-\boldsymbol{\mathcal{F}}(\boldsymbol{\psi})
	\big\|_{2}^{2}
	\right).
	\label{eq:map_likelihood}
\end{align}
Substituting~\eqref{eq:map_likelihood} into~\eqref{eq:map_general_compact} yields the equivalent MAP optimization:
\begin{align}
	\hat{\boldsymbol{\psi}}_{\mathrm{sat}}
	=\arg\min_{\boldsymbol{\psi}}
	\Big[
	\frac{1}{2\sigma_{n}^{2}}
	\big\|
	\mathbf{P}-\boldsymbol{\mathcal{F}}(\boldsymbol{\psi})
	\big\|_{2}^{2}
	-\ln p(\boldsymbol{\psi})
	\Big].
	\label{eq:map_cost_compact}
\end{align}
For a uniform prior distribution over the acquisition region, $p(\boldsymbol{\psi})=\mathrm{const.}$, and the MAP estimate simplifies to the Maximum Likelihood (ML) solution:
\begin{align}
	\hat{\boldsymbol{\psi}}_{\mathrm{sat}}^{\mathrm{ML}}
	=\arg\min_{\boldsymbol{\psi}}
	\big\|
	\mathbf{P}-\boldsymbol{\mathcal{F}}(\boldsymbol{\psi})
	\big\|_{2}^{2}.
	\label{eq:ml_est_compact}
\end{align}

The mapping $\boldsymbol{\mathcal{F}}(\boldsymbol{\psi})$ in~\eqref{eq:map_obs_compact} is highly nonlinear, coupled, and non-invertible due to the combined effects of beam divergence, turbulence, and reflection geometry. 
Hence, direct numerical or analytical minimization of~\eqref{eq:map_cost_compact} is computationally intractable, motivating a data-driven inference framework to approximate the inverse mapping $\mathbf{P}\!\mapsto\!\boldsymbol{\psi}_{\mathrm{sat}}$.

\subsection{Learning-Based Position Estimation}\label{sec:learning}

\subsubsection{Justification}
Given the MAP program in \eqref{eq:map_cost_compact} and the observation model \eqref{eq:map_obs_compact}, a tractable solver is impeded by: 
(i) the nonconvex, non-invertible mapping $\boldsymbol{\mathcal{F}}(\cdot)$ induced by the super-Gaussian footprints \eqref{eq:beam_pattern_xy} and geometric overlap \eqref{eq:hpt_approx_final} (with turbulence-dependent fading embedded in $\boldsymbol{\mathcal{F}}$); 
(ii) background-induced fluctuations and model mismatch; and 
(iii) unknown or time-varying nuisance parameters (e.g., effective divergences, reflectivity, partial occlusion). 
Hence, direct minimization of \eqref{eq:map_cost_compact} is generally intractable.

\subsubsection{Learning-Based Inverse Mapping}
We approximate the inverse of \eqref{eq:map_obs_compact} by a deep regressor
\begin{align}
	f_{\boldsymbol{\theta}}:\ \mathbb{R}^{N_L+N_T}\!\rightarrow\!\mathbb{R}^{2}, 
	\qquad 
	\hat{\boldsymbol{\psi}}_{\mathrm{sat}}=f_{\boldsymbol{\theta}}(\mathbf{P}), 
	\label{eq:learned_inverse_final}
\end{align}
where $\boldsymbol{\theta}$ denotes trainable parameters. 
With sufficient training data, $f_{\boldsymbol{\theta}}$ (e.g., an MLP or Transformer network) can approximate the MAP or MMSE estimator of $\boldsymbol{\psi}_{\mathrm{sat}}$.

\subsubsection{Training Objective (AWGN-Consistent)}
Let $\mathcal{D}=\{(\mathbf{P}_i,\boldsymbol{\psi}_i)\}_{i=1}^{N}$ denote a supervised dataset generated from \eqref{eq:map_obs_compact}, where $\boldsymbol{\psi}_i=[x_i,\,y_i]^{\!\top}$. 
Under the AWGN assumption used in \eqref{eq:map_likelihood}–\eqref{eq:map_cost_compact}, empirical risk minimization with $\ell_2$ loss provides a consistent surrogate:
\begin{align}
	\hat{\boldsymbol{\theta}}
	= \arg\min_{\boldsymbol{\theta}}
	\frac{1}{N}\sum_{i=1}^{N}
	\big\|f_{\boldsymbol{\theta}}(\mathbf{P}_i)-\boldsymbol{\psi}_i\big\|_2^2
	+\lambda\,\mathcal{R}(\boldsymbol{\theta}),
	\label{eq:erm_awgn}
\end{align}
where $\mathcal{R}$ denotes a weight-decay or convex regularizer and $\lambda\!\ge\!0$. 
Additionally, a soft feasibility prior consistent with the acquisition disk can be imposed on the outputs:
\begin{align}
	\mathcal{R}_{\text{out}}
	=\beta\,\frac{1}{N}\sum_{i=1}^{N}
	\big\|\Pi_{\mathcal{B}(0,R_{\mathrm{u}})}\!\big(f_{\boldsymbol{\theta}}(\mathbf{P}_i)\big)
	- f_{\boldsymbol{\theta}}(\mathbf{P}_i)\big\|_2^{2},
	\label{eq:soft_prior}
\end{align}
where $\beta\!\ge\!0$, and $\Pi_{\mathcal{B}(0,R_{\mathrm{u}})}$ denotes the projection operator onto the radius-$R_{\mathrm{u}}$ disk.

\subsubsection{Physics-Consistent Featureization}
To incorporate the scanning structure into the learning model, beam-wise standardization and positional encoding of scan indices are applied:
\begin{align}
	\tilde{\mathbf{P}}
	=\mathrm{Norm}(\mathbf{P})\ \oplus\ \boldsymbol{\phi}_{\mathrm{scan}},
	\qquad 
	\hat{\boldsymbol{\psi}}_{\mathrm{sat}}
	=f_{\boldsymbol{\theta}}(\tilde{\mathbf{P}}),
	\label{eq:feat_embed}
\end{align}
where $\boldsymbol{\phi}_{\mathrm{scan}}$ represents positional encodings reflecting the $(x$-sweep$,\ y$-sweep$)$ order (e.g., sinusoidal or one-hot index embedding).

\subsubsection{Inference and Complexity}
At inference, position estimation is achieved via a single forward pass with an optional projection:
\begin{align}
	\hat{\boldsymbol{\psi}}_{\mathrm{sat}}
	=\Pi_{\mathcal{B}(0,R_{\mathrm{u}})}\!\big(f_{\boldsymbol{\theta}}(\tilde{\mathbf{P}})\big).
	\label{eq:inference_project}
\end{align}
For an $L$-layer MLP with layer widths $\{w_\ell\}_{\ell=1}^{L}$ and two outputs, the computational cost of inference scales as
\begin{align}
	\mathcal{C}_{\mathrm{infer}}
	=\mathcal{O}\!\Big((N_L\!+\!N_T)\,w_1+\sum_{\ell=1}^{L-1} w_\ell w_{\ell+1}+2\,w_L\Big),
	\label{eq:complexity_final}
\end{align}
which enables millisecond-scale updates within each $T_{\mathrm{scan}}$ interval.

\begin{table*}[t]
	\centering
	\caption{Comparison of positioning performance for the MAP, learning-driven, and literature-based methods under four representative test cases (all values in meters, with $T_{\mathrm{scan}}$ given in milliseconds).}
	\label{tab:pos_estimation_summary}
	\renewcommand{\arraystretch}{1.2}
	\setlength{\tabcolsep}{3.5pt}
	\scriptsize
	\begin{tabular}{|l|cccc|cccc|cccc|cccc|}
		\hline
		\multirow{2}{*}{\textbf{Method}} &
		\multicolumn{4}{c|}{\textbf{Test 1}} &
		\multicolumn{4}{c|}{\textbf{Test 2}} &
		\multicolumn{4}{c|}{\textbf{Test 3}} &
		\multicolumn{4}{c|}{\textbf{Test 4}} \\
		\cline{2-17}
		& $x_{\mathrm{sat}}$ & $y_{\mathrm{sat}}$ & $e$ & $T_{\mathrm{scan}}$ &
		$x_{\mathrm{sat}}$ & $y_{\mathrm{sat}}$ & $e$ & $T_{\mathrm{scan}}$ &
		$x_{\mathrm{sat}}$ & $y_{\mathrm{sat}}$ & $e$ & $T_{\mathrm{scan}}$ &
		$x_{\mathrm{sat}}$ & $y_{\mathrm{sat}}$ & $e$ & $T_{\mathrm{scan}}$ \\
		\hline
		Real (True) &
		$-452.4$ & $215.5$ & -- & -- &
		$-181.8$ & $314.9$ & -- & -- &
		$-733.4$ & $-269.4$ & -- & -- &
		$489.6$ & $-524.8$ & -- & -- \\
		\hline
		MAP &
		$-448.9$ & $212.7$ & $6.6$ & $1.0$ &
		$-187.3$ & $309.7$ & $8.8$ & $1.0$ &
		$-738.5$ & $-272.2$ & $10.3$ & $1.0$ &
		$493.8$ & $-528.9$ & $7.9$ & $1.0$ \\
		\hline
		Learning-based &
		$-456.0$ & $211.4$ & $7.0$ & $1.0$ &
		$-187.7$ & $307.1$ & $14.5$ & $1.0$ &
		$-739.0$ & $-263.7$ & $12.2$ & $1.0$ &
		$488.2$ & $-531.0$ & $9.4$ & $1.0$ \\
		\hline
		Literature-based &
		$-445.2$ & $207.6$ & $14.8$ & $1.4$ &
		$-177.4$ & $322.4$ & $11.3$ & $0.5$ &
		$-727.5$ & $-266.9$ & $7.7$ & $3.5$ &
		$494.5$ & $-520.1$ & $10.2$ & $2.6$ \\
		\hline
	\end{tabular}
\end{table*}

\begin{figure*}
	\centering
	\subfloat[] {\includegraphics[width=3.4 in, height= 1.8 in]{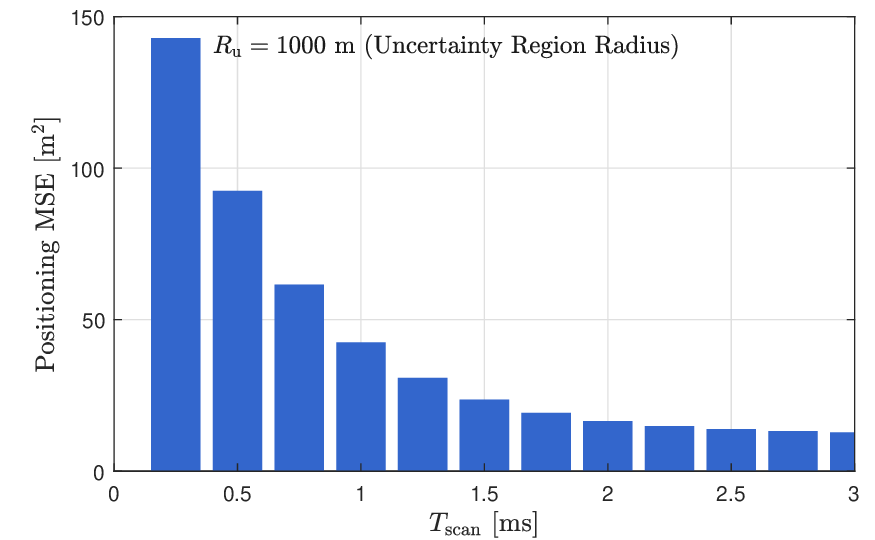}
		\label{ck1}
	}
	\hfill
	\subfloat[] {\includegraphics[width=3.4 in, height= 1.8 in]{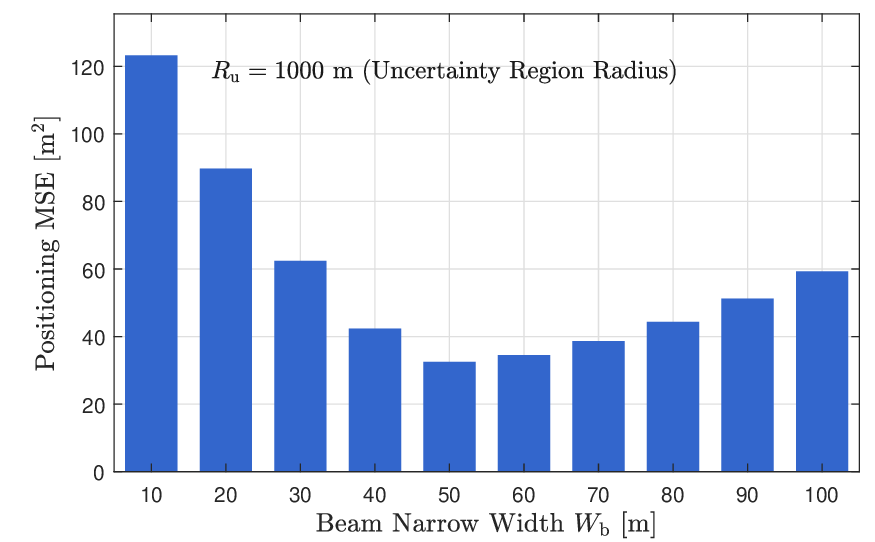}
		\label{ck2}
	}
	\caption{Parametric performance analysis of the proposed learning-driven dual-line laser scanning system for LEO positioning. 
		(a)~Positioning MSE versus total scanning time $T_{\mathrm{scan}}$: the MSE decreases rapidly with longer scan duration and saturates near $T_{\mathrm{scan}}\!\approx\!2~\mathrm{ms}$, indicating diminishing returns beyond this point. 
		(b)~Positioning MSE versus beam narrow width $W_{\mathrm{b}}$: the minimum $\mathrm{MSE}_{\mathrm{pos}}\!\approx\!5.7^2~\mathrm{m}^2$ occurs at $W_{\mathrm{b}}\!=\!50~\mathrm{m}$, while both narrower ($<50~\mathrm{m}$) and wider ($>50~\mathrm{m}$) beams lead to degraded accuracy. 
		Results are generated for an uncertainty region radius of $R_{\mathrm{u}}\!=\!1000~\mathrm{m}$.}
	
	\label{ck}
\end{figure*}

\section{Performance Evaluation and Discussion}
To evaluate the performance of the proposed learning-driven dual-line laser scanning framework, 
a physics-based simulator was developed following the validated analytical models presented in Sections~II–III, 
which are consistent with experimentally verified formulations reported in the literature~\cite{AndrewsBook,parent1992propagation,10553228,10681506}.
The main simulation parameters are as follows: 
the link distance is $Z_q=10^6~\mathrm{m}$ (typical LEO altitude), 
the ambiguity region radius is $R_{\mathrm{u}}=1000~\mathrm{m}$, 
and each beam transmits $P_t=5~\mathrm{W}$ of optical power with overall efficiency $\eta_{\mathrm{MRR}}=0.8$. 
The return-beam divergence is set to $\theta_{\mathrm{div}}=30~\mu\mathrm{rad}$, 
the receiver aperture radius to $d_r=0.25~\mathrm{m}$, 
and the effective MRR array area to $A_{\mathrm{MRR}}=0.04~\mathrm{m}^2$.
The atmospheric turbulence is modeled by a Gamma–Gamma distribution with a ground-level refractive-index structure parameter of $C_{n}^{2}(0)=10^{-13}~\mathrm{m}^{-2/3}$, 
obtained by integrating along the satellite’s oblique propagation path, consistent with~\cite{10681506,AndrewsBook}.
The dual-line scanner employs $N_\mathrm{L}=N_\mathrm{T}=40$ discrete scan positions, 
and unless otherwise stated, the total scanning period is $T_{\mathrm{scan}}=1~\mathrm{ms}$, 
corresponding to a dwell time of $\tau_{\mathrm{hit}}=T_{\mathrm{scan}}/N_\mathrm{T}$ per step, 
which represent typical values for MRR-based LEO optical links~\cite{AndrewsBook,10553228,10681506}. 
The line-beam divergence is adjusted such that its longitudinal coverage spans $2000~\mathrm{m}$ over the full scanning region, 
while its default narrow width is $W_{\mathrm{b}}=40~\mathrm{m}$, 
whose impact on positioning accuracy is analyzed in detail in Fig.~\ref{ck2}.

Synthetic energy–position datasets were generated from the physical model.
Input features were standardized and normalized, and the satellite positions $\mathbf{x}_{\mathrm{sat}}=[x_{\mathrm{sat}},y_{\mathrm{sat}}]^{\mathsf{T}}$ were scaled by $R_{\mathrm{u}}$.
An MLP with two hidden layers (512–256) and ReLU activations was trained using the Adam optimizer ($\eta_0=5\times10^{-4}$, batch size 512) with 15\% validation split and early stopping.
The network outputs normalized coordinates $\hat{x},\hat{y}$ that are rescaled back to meters at inference.

Table~\ref{tab:pos_estimation_summary} summarizes four representative test cases obtained from the proposed simulator. 
As expected, the MAP estimator achieves the lowest positioning errors, typically in the range of 5–10~m, 
but at the cost of significantly higher computational complexity. 
The proposed learning-driven dual-line method provides a close approximation to the MAP results 
while requiring only a single forward pass during inference, thus enabling real-time operation. 
The literature baseline~\cite{10553228} follows a two-stage procedure, 
where the ambiguity region is first sensed through a random scanning process using conventional Gaussian beams. 
After this stochastic sensing phase, a separate positioning stage is performed using three Gaussian beams to triangulate the satellite’s location. 
Because the sensing duration is random, this method exhibits non-deterministic latency and a variable positioning accuracy. 
As observed in Table~\ref{tab:pos_estimation_summary}, its scanning time varies from $0.5$ to $3.5~\mathrm{ms}$ across different test cases, 
leading to positioning errors between $7.7$ and $14.8~\mathrm{m}$. 
In contrast, both the MAP and learning-driven methods operate deterministically with $T_{\mathrm{scan}}=1~\mathrm{ms}$. 
The MAP estimator consistently achieves the smallest errors (e.g., $6.6~\mathrm{m}$ in Test~1 and $7.9~\mathrm{m}$ in Test~4), 
while the proposed learning-driven dual-line approach provides nearly similar accuracy (e.g., $7.0$ and $9.4~\mathrm{m}$ for the same tests) 
with negligible computational cost and identical scanning time. 
These results confirm that the proposed learning-based method achieves a favorable balance between precision, stability, and latency compared to both the MAP and literature-based schemes.

Figures~\ref{ck1} and~\ref{ck2} further illustrate the sensitivity of the proposed learning-based system to key design parameters. 
Figure~\ref{ck1} shows that the positioning mean squared error (MSE) decreases rapidly as the total scanning time $T_{\mathrm{scan}}$ increases from $0.25$ to about $2~\mathrm{ms}$. 
Beyond this point, the MSE converges to a nearly constant level of approximately $3^2~\mathrm{m}^2$, 
demonstrating that longer scanning times provide little additional benefit while increasing system latency. 
This confirms that a scan time of $1$--$2~\mathrm{ms}$ offers an effective trade-off between accuracy and real-time operation.

In contrast, Figure~\ref{ck2} analyzes the influence of the beam narrow width $W_{\mathrm{b}}$ on the positioning accuracy. 
The minimum positioning error is achieved at $W_{\mathrm{b}}=50~\mathrm{m}$, where $\mathrm{MSE}_{\mathrm{pos}}\approx5.7^2~\mathrm{m}^2$. 
For narrower beams ($W_{\mathrm{b}}<40~\mathrm{m}$), the MSE increases due to stronger sensitivity to turbulence and pointing jitter, 
while for wider beams ($W_{\mathrm{b}}>80~\mathrm{m}$), the MSE also rises because of reduced optical intensity and lower spatial resolution. 
At $W_{\mathrm{b}}=10~\mathrm{m}$, the MSE reaches approximately $11.1^2~\mathrm{m}^2$, 
and at $W_{\mathrm{b}}=100~\mathrm{m}$ it increases to about $7.7^2~\mathrm{m}^2$. 
These results highlight that the optimal beam width around $50~\mathrm{m}$ ensures the best compromise between energy concentration and spatial coverage for an ambiguity radius of $R_{\mathrm{u}}=1000~\mathrm{m}$.

Overall, the presented results validate that the learning-driven dual-line scanning system achieves 
near-MAP accuracy within a deterministic 1--2~ms scanning window, 
making it a viable solution for next-generation LEO optical positioning and tracking applications.

\bibliographystyle{IEEEtran}
\bibliography{IEEEabrv,myref}

\end{document}